\documentclass[twocolumn,showpacs,amsmath,amssymb,prd]{revtex4}

\usepackage{graphicx}
\usepackage{dcolumn}
\usepackage{bm}

\begin{document}
\title{Analysis of solar neutrino problem by means of N\"{o}tzold and Nakagawa's approach including the interference term\\- Hyperbolic-tangent profile for electron density in the sun and exact solution -}

\author{M. Kaneyama}
\email[Email address: ]{kaneyama@azusa.shinshu-u.ac.jp}
\author{M. Biyajima}
\affiliation{Department of physics, Faculty of Science, Shinshu University, 390-8621 Matsumoto, Japan}

\date{\today}

\begin{abstract}
Using an exact solution with the hyperbolic-tangent profile for the electron density in the sun, which is developed by N\"{o}tzold and later modified by Nakagawa, we have analyzed the solar neutrino problem. An interference term in their approach is correctly taken into account. Combining the hyperbolic-tangent profile with the BP2000, we obtain a phenomenological fitting in the analytic form. Combining recent observed results for survival probability $P(\nu_e \rightarrow \nu_e)$ by the SNO, SK, SAGE, Gallex, GNO and Homestake Collaborations, we obtain a large mixing angle (LMA) whose figure is looking like a shoulder.
\end{abstract}


\maketitle

\section{Introduction}
One of the most interesting subjects in elementary particle physics is the solar neutrino problem. As a possible explanation for it, the MSW mechanism seems the best solution \cite{wolfen,MS,bethe}. Very recently the SNO Collaboration has reported that the survival probability of $ \nu_e $ from the sun is $ P(\nu_e \rightarrow \nu_e)=0.348\pm 0.03$, and the large mixing angle (LMA), $ \Delta m^2=m_2^2-m_1^2=0.35\times 10^{-4}(\text{eV}^2)$ and $ \theta=32^\circ $, is favor for the explanation of the global measurements\cite{sno1,sno2,sno3} ($m_1$, $m_2$ are masses of the neutrino mass eigenstate $\nu_1$, $\nu_2$, respectively. Hereafter $m_2>m_1$ is assumed. $\theta$ is the mixing angle defined by $\nu_e=\cos\theta\,\nu_1+\sin\theta\,\nu_2$ and $\nu_\mu=-\sin\theta\,\nu_1+\cos\theta\,\nu_2$. ).

\begin{figure}[h]
    \includegraphics[keepaspectratio=true,height=85mm]{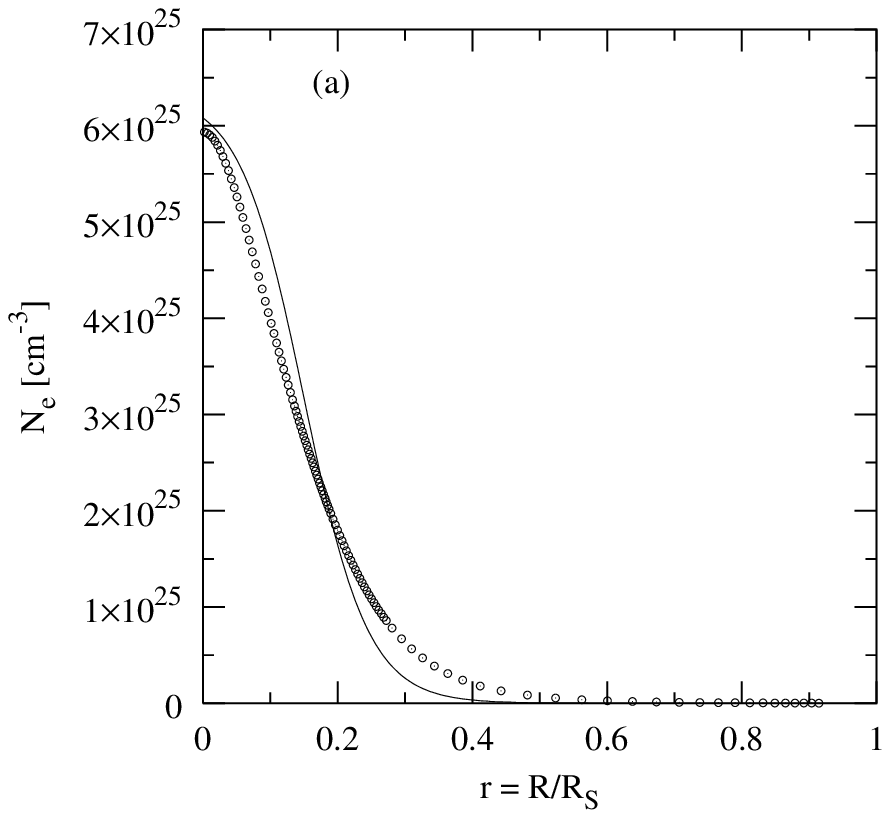}
    \includegraphics[keepaspectratio=true,height=85mm]{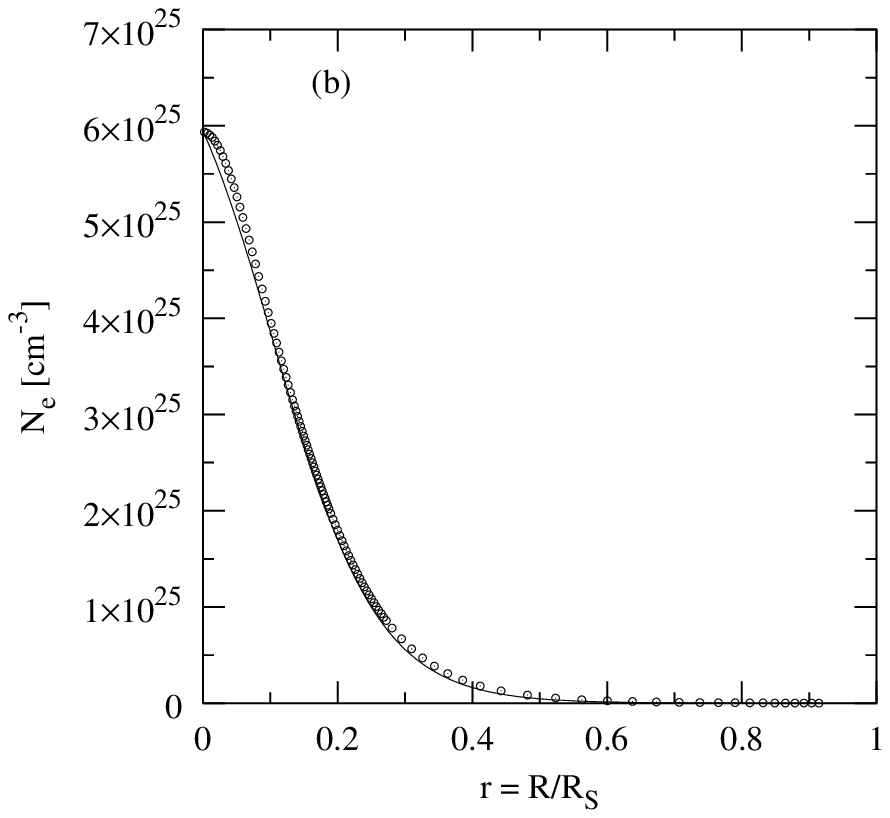}
    \caption{(a) The electron number density from SSM1988 in Ref.\cite{ssm88}, which is fitted by Eq.\eqref{eq:noepotential}. See Fig.2 of Ref.\cite{Noetzold}. (b) $N_e$ is fitted by Eq.\eqref{eq:nakapotential}. See Ref.\cite{nakagawa}.}
  \label{fig:ssm88}
\end{figure}

On the other hand in 1988 N\"{o}tzold proposed an exact solution for the solar neutrino oscillation, using the hyperbolic-tangent profile for the density of electrons defined by $N_e$ in the sun\cite{Noetzold} :

\begin{equation}
N_e=N_0[1-\tanh\frac{r}{l}], \label{eq:noepotential}
\end{equation}
where $N_0=V\,/\,(R_S\sqrt{2}G_F)$, $R_S$ and $G_F$ denote the radius of sun $R_S=6.96\times 10^6$km and the Fermi coupling constant $G_F=8.917\times 10^{-8}(\text{GeV}\cdot\text{fm}^3)$. Moreover $V$, $r$ and $l$ denote the magnitude of the density at the center of the sun, $r=R/R_S$ and $l=l/R_S$, respectively. N\"{o}tzold has assumed $r \rightarrow r+r_0=r-\infty$, a priori (See Fig.\ref{fig:ssm88}(a)).
  
Later, Nakagawa has proposed a modified expression for $N_e$ as
\begin{equation}
N_e=N_0[1-\tanh\frac{r+r_0}{l}], \label{eq:nakapotential}
\end{equation}
where $r_0$ is an adjusted parameter which reproduces the $N_e$ in better way than Eq.\eqref{eq:noepotential}.
Nakagawa has stressed that the result by N\"{o}tzold, the magnitude of $\Delta m^2/E_\nu$, is reduced about 26\% in the figure of $\Delta m^2/E_\nu$ vs. $\sin^2\theta$, as Eq.\eqref{eq:nakapotential} is used for the electron density in the sun\cite{nakagawa}. His result with the standard solar model (SSM) in 1988\cite{ssm88} is shown in Fig.\ref{fig:ssm88}(b) ($V=1.718\times 10^4$, $l=0.155$, $r_0=-0.105$ for SSM in 1988).

\begin{figure}[h]
  \begin{center}
    \includegraphics[keepaspectratio=true,height=85mm]{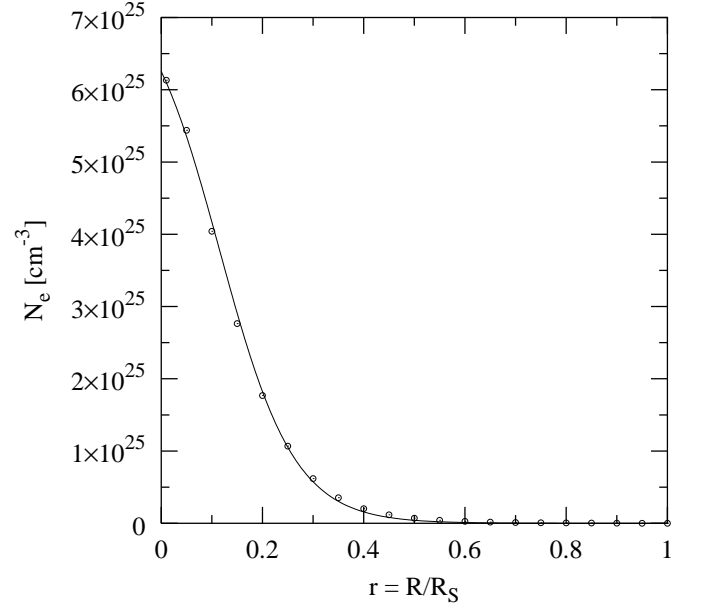}
  \end{center}
  \caption{The electron number density from BP2000, which is fitted by             Eq.\eqref{eq:nakapotential}.}
  \label{fig:bp2000}
\end{figure}

In this report we would like to consider the BP2000\cite{bp2000} with Eq.\eqref{eq:nakapotential} and the solar neutrino problem based on N\"{o}tzold-Nakagawa's approach including the interference term. Our phenomenological fitting is shown in Fig.\ref{fig:bp2000} ($V=1.740\times 10^4$, $l=0.148$, $r_0=-0.115$ for BP2000).

In 2nd section, their theoretical formulas a la Nakagawa are explained. The explicit interference term is shown. In 3rd section, we estimate the allowed regions ($\Delta m^2 $ and mixing angle) for survival probabilities observed by the SNO, SK, SAGE, GNO, Gallex and Homestake Collaborations\cite{sno1,sno2,sno3,sk1,sk2,sage,gno,home98}. In 4th section, concluding remarks are presented.

\section{\label{formula}Theoretical formulas with the hyperbolic-tangent for $N_e$}
To explain theoretical formulas first derived by N\"{o}tzold and later modified by Nakagawa, we briefly describe their framework. The coupled Schr\"{o}dinger equations for the solar neutrino oscillation are given as 
\begin{align}
i\frac{d\nu_e}{dt}= \biggl( \frac{m_1^2\cos^2\theta+m_2^2\sin^2\theta}{2E}+&\sqrt{2}G_FN_e \biggr) \nu_e\notag \\
+&\frac{\Delta m^2}{2E}\sin\theta\cos\theta\nu_\mu,
\end{align}
\begin{equation}
i\frac{d\nu_\mu}{dt}=\frac{\Delta m^2}{2E}\sin\theta\cos\theta\nu_e+\frac{m_1^2\sin^2\theta+m_2^2\cos^2\theta}{2E}\nu_\mu.
\end{equation}
Introducing the following notations, 
$ct=R_S\,r$, $M_0^2=R_S(m_2^2+m_1^2)/(4E)$, $R_S\Delta m^2/(2E)=x/B\equiv x_0$
with $B\equiv E(\text{MeV})/\Delta m^2(\text{eV}^2)$ and $x=1.764\times 10^9$, we use the following abbreviations 
\begin{equation}
G=\frac{x_0\cos 2\theta}{2},
\end{equation}
\begin{equation}
H=\frac{x_0\sin 2\theta}{2}.
\end{equation}
Moreover, exchanging the variable from $\nu_e$ to $v(r)$, 
\begin{equation}
\nu_e(r)=v(r)\exp\left\{-i\int^{r}_{0}[M_0^2+\frac{1}{2}VU(r)]dr\right\},
\end{equation}
we have the following differential equation as 

\begin{equation}
\frac{d^2v}{dr^2}+\left[i\frac{1}{2}VU(r)^\prime+\{\frac{1}{2}VU(r)-G\}^2+H^2\right]v=0\,,\label{eq:2d}
\end{equation}
where $U(r)=1-\tanh[(r+r_0)/l\,]$. Changing the variable $r$ to $y$ and assuming the factorized form for $v(r)$, 
\begin{equation}
y=\cfrac{1}{1+\exp\left(-2\cfrac{r+r_0}{l}\right)}\; ,
\end{equation}
\begin{equation}
U(r)=2(1-y),
\end{equation}
\begin{equation}
v(r)=(1-y)^\mu y^\nu f(y),
\end{equation}
we obtain the following hyper-geometric equation
\begin{equation}
y(1-y)\frac{d^2f}{dy^2}+\{c-(a+b+1)y\}\frac{df}{dy}-abf=0,\label{eq:h.geo}
\end{equation}
where
\begin{equation}
a=\mu+\nu+\lambda,
\end{equation}
\begin{equation}
b=\mu+\nu+1-\lambda,
\end{equation}
\begin{equation}
c=2\nu+1,
\end{equation}
\begin{equation}
\mu=i\frac{l}{2}\sqrt{G^2+H^2}\, ,
\end{equation}
\begin{equation}
\nu=i\frac{l}{2}\sqrt{(V-G)^2+H^2}\, ,
\end{equation}
\begin{equation}
\lambda=i\frac{l}{2}V\,.
\end{equation}
The general solution of Eq.\eqref{eq:h.geo} is expressed by two independent hyper-geometrical functions. $F(a,b;c;y)_{-}$ is expressed by replacing $\nu\rightarrow -\nu$ in $a$, $b$, $c$ of $F(a,b;c;y)_{+}$.
\begin{equation}
f(y)=C_1F(a,b;c;y)_{+}+C_2y^{-2\nu}F(a,b;c;y)_{-},
\end{equation}
where $C_1$ and $C_2$ are the integral constants. Using an ordinary procedure with the initial condition, $\nu_e(y=y_0)=1$ and $\nu_\mu(y=y_0)=0$, we obtain the following formula for the survival probability
\begin{gather}
\langle P_{\nu_e\rightarrow\nu_e}\rangle=P_1\cos^2\theta+(1-P_1)\sin^2\theta \notag \\
-\sqrt{P_c(1-P_c)}\cos2\theta_m\sin2\theta\cos(2.54\frac{\Delta m^2}{E}L+\delta)\;.\label{eq:s.probability}
\end{gather}
where
\begin{equation}
P_1=P_c\sin^2\theta_m+(1-P_c)\cos^2\theta_m,
\end{equation}
\begin{equation}
P_c=\frac{\cosh(\pi lV)-\cosh[\pi l(\Delta p-\Delta q)]}{\cosh[\pi l(\Delta p+\Delta q)]-\cosh[\pi l(\Delta p-\Delta q)]},
\end{equation}
\begin{equation}
\cos2\theta_m=\frac{G-V(1-y_0)}{\sqrt{[G-V(1-y_0)]^2+H^2}},
\end{equation}
\begin{equation}
\Delta p=\frac{R_S}{4E}\Delta m^2,
\end{equation}
\begin{equation}
\Delta q=\sqrt{V^2-2V\Delta p\cos2\theta+\Delta p^2},
\end{equation}
\begin{equation}
y_0=\cfrac{1}{1+\exp(-2\cfrac{r_0}{l})}\; .
\end{equation}
In Refs.\cite{Noetzold} \cite{nakagawa} the third term of the right hand side, the interference term, in Eq.\eqref{eq:s.probability} is assumed to be zero because of the oscillation. We have obtained following formula, after integration,
\begin{gather}
-\frac{R_S}{10.16\times10^3\Delta L\Delta p}\sqrt{P_c(1-P_c)}\cos2\theta_m\sin2\theta \notag \\
\times\sin(\frac{5.08\times10^3}{R_S}\Delta L\Delta p)\cos\left[\frac{5.08\times10^3}{R_S}(L_2+L_1)\Delta p\right]\;,\label{3term.s.probability}
\end{gather}
where $L_1$ and $L_2$ are the minimum distance from the sun to earth and the maximum one, respectively, and
\begin{equation}
\Delta L=L_2-L_1\;.
\end{equation}

\section{\label{analyses}Analysis of solar neutrino problem with empirical values by means of Eq.(20)}
The data on the solar neutrino, the survival probability $P(\nu_e\rightarrow\nu_e)$, reported by the SNO, Gallex, GNO, SAGE, Homestake Collaborations are shown in Table \ref{table:data}.

\begin{table}[htbp]
 \caption{\label{table:data} Data on the solar neutrino\cite{sno3,sage,gno,home98}. $\langle E_\nu\rangle$ is the average neutrino energy. }
 \begin{center}
  \begin{tabular}{|c|c|c|}
    \hline
      Exp & $\langle E_\nu\rangle$ &$P(\nu_e\rightarrow\nu_e)$  \\
    \hline
      SNO & $\sim8$ MeV  & $0.348\pm0.029$ \\
    \hline
      SAGE& $\sim0.8$ MeV & $0.54\pm0.06$   \\
    \hline
      Gallex + GNO & $\sim0.8$ MeV &$0.56\pm0.07$    \\
    \hline
      Homestake & $\sim8$ MeV & $0.34\pm0.03$   \\
    \hline
  \end{tabular}
 \end{center}
\end{table}

\begin{figure}[h]
    \includegraphics[keepaspectratio=true,height=85mm]{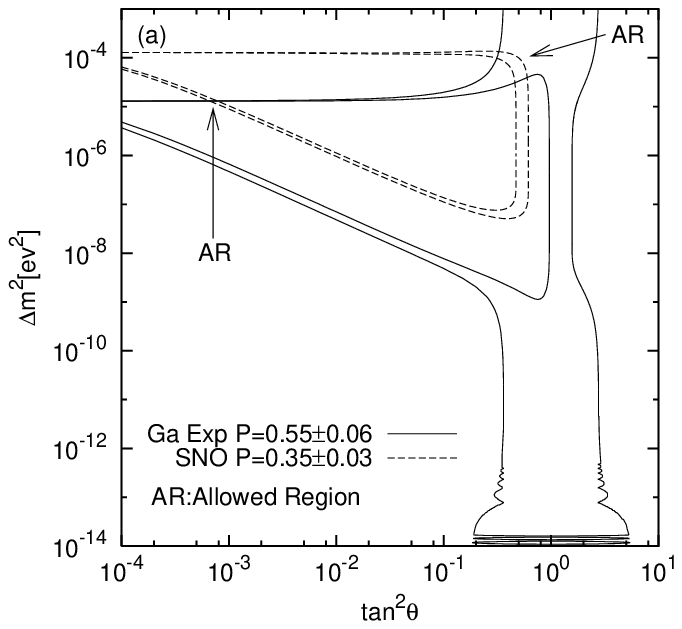}  
    \includegraphics[keepaspectratio=true,height=85mm]{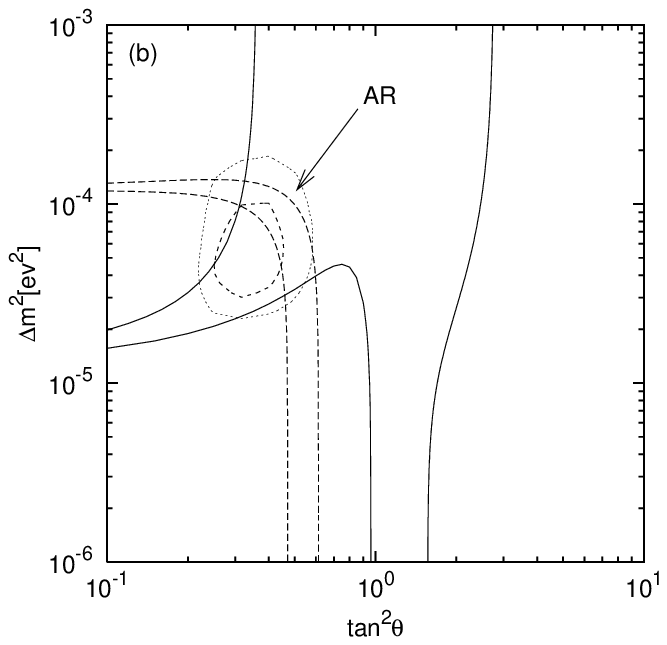}             
  \caption{(a) $\Delta m^2$ vs. $\tan^2\theta$. Long dashed and solid lines are obtained from empirical values by SNO Collaboration ($P(\nu_e\rightarrow\nu_e)=0.348\pm 0.03)$, and those by Ga experiment($P(\nu_e\rightarrow\nu_e)=0.55\pm 0.06$) and assuming $\langle E_\nu\rangle=8\text{MeV}$ and $\langle E_\nu\rangle=0.8\text{MeV}$, respectively. (b) Enlarged figure of (a). The dashed circle with $\text{C.L.}=90\%$ and dotted circle with $\text{C.L.}=99.73\%$ are given in Ref.\cite{sno1}. The SMA is excluded by observations by the SNO and SK Collaboration. AR is an abbreviation for allowed region.}
  
    \label{fig:region}
\end{figure}

Using empirical values in Table \ref{table:data} and Eq.\eqref{eq:s.probability} with $\langle E_\nu \rangle =8\text{MeV}$ and $\langle E_\nu \rangle =0.8\text{MeV}$, we obtain two allowed regions (AR) which are shown in Fig.\ref{fig:region}(a). The small mixing angle (SMA) region is located in $(\tan^2\theta,\,\Delta m^2)=(0.8\times10^{-3},\,10^{-5})$ and the large mixing angle (LMA) region is located in $(0.5,\,10^{-4}\sim 4\times10^{-5})$.
Combining the global measurement (including the day-night effect by the SNO Collaboration and by the SK Collaboration) and our result,  we obtain Fig.\ref{fig:region}(b). The shoulder-like region nearly coincides with the large mixing angle (LMA) with C.L.99.73\%. (See Ref.\cite{sno1})

As is seen in Fig.\ref{fig:region}, our results based on Eq.\eqref{eq:nakapotential} 
and Eq.\eqref{eq:s.probability} is partially consistent with the LMA solutions.

\section{\label{remarks}Concluding remarks}
\begin{enumerate}
\item 
As is seen in Fig.\ref{fig:bp2000}, Eq.\eqref{eq:nakapotential} is useful to reproduce the electron density based on BP2000.
\item
Using Eq.\eqref{eq:s.probability}, we have examined  the AR expressed by ($\Delta m^2$, $\tan^2\theta$). In the present calculation we have added the third  term Eq.\eqref{3term.s.probability} to N\"{o}tzold and Nakagawa's formula. As compared with result of Refs.\cite{Noetzold}\cite{nakagawa}, we have additional AR: The contribution is seen in the most right curve and the bottom region  with oscillations near ($\Delta m^2$, $\tan^2\theta$) $=$ ($0.2\sim5$, $1\times10^{-14}\sim1\times10^{-13}$).The main reason is attributed to Eq.\eqref{3term.s.probability}. Moreover, we can compare our results with the exponential profile, for example, see Fig.2 of Ref.\cite{murayama}. A similar allowed region ($0.1\lesssim \tan^2\theta\lesssim1.0$) is observed.

\item 
From the empirical results in Table \ref{table:data} and Eq.\eqref{eq:s.probability}, we have obtained two AR's, the LMA and SMA. Combining the AR's reported by the SNO, the SK Collaborations and our Fig.\ref{fig:region}(a), we have obtained Fig.\ref{fig:region}(b). This is fairly well consistent with the LMA reported by the SNO Collaboration.
 
\item 
Moreover, we have to add the following fact; For the figure of ($\sin^2\theta$, $\Delta m^2/E_\nu$) with the BP2000, the result of calculation a la Ref.\cite{Noetzold}, the magnitude of $\Delta m^2/E_\nu$, is about 20\% larger than that of present calculation using values reported by the SNO Collaboration\cite{sno1}.
\end{enumerate}

\begin{acknowledgments}
The authors would like to thank their colleagues for their assistance, in particular T. Mizoguchi, and useful communication with T.Kaneko and H.Nunokawa.
\end{acknowledgments}

\end{document}